\documentstyle[twoside,fleqn,espcrc2,epsf]{article}
\def\fig#1#2#3{\epsfxsize=#3truein
\vskip -0.3 truein
\centerline{\epsffile{fig#1.eps}}
\centerline{\vbox{{\bf \noindent Figure #1.} #2}}
\bigskip}

\def\figsizeA{2.1}
\def\figsizeB{2.0}
\def\figsizeC{2.2}
\def\pbp{\langle\overline{\psi}\psi\rangle}

\def\spose#1{\hbox to 0pt{#1\hss}}
\def\ltapprox{\mathrel{\spose{\lower 3pt\hbox{$\mathchar"218$}}
 \raise 2.0pt\hbox{$\mathchar"13C$}}}
\def\gtapprox{\mathrel{\spose{\lower 3pt\hbox{$\mathchar"218$}}
 \raise 2.0pt\hbox{$\mathchar"13E$}}}
\def\inapprox{\mathrel{\spose{\lower 3pt\hbox{$\mathchar"218$}}
 \raise 2.0pt\hbox{$\mathchar"232$}}}

\def\one{
$8^4 \times 4$, $L_s=12$, $m_f=0.1$, and $m_0=1.9$.  $|W|$ (crosses)
and $\pbp$ (diamonds).}
\def\two{ $8^3\times 4$ $\beta=5.20$ (squares), $8^3 \times 4$
$\beta=5.45$ (diamonds), $16^3 \times 4$ $\beta=5.45$ (crosses), with
$L_s=16$ and $m_0=1.9$.  The stars are the $m_f=0$ linearly
extrapolated values.}
\def\three{$8^4 \times 4$, $m_f=0.1$ and $m_0=1.9$.  $\beta=5.20$
(crosses), $\beta=5.45$ (diamonds).}
\def\four{$8^4 \times 4$, $L_s=12$ and $m_f=0.1$. 
$\beta_c$ from $|W|$ (crosses) and from $\pbp$ (diamonds).}
\def\five{$16^4 \times 4$, $L_s=16$, $\beta=5.45$, and $m_0=1.9$.  The
fit is to $c_0 + c_2 m_f^2$ and the star is the $m_f=0$ extrapolated
value.}
\newcommand{\AmS}{{\protect\the\textfont2
  A\kern-.1667em\lower.5ex\hbox{M}\kern-.125emS}}

\title{
\vspace{-3.5cm}
\begin{flushright}
{\normalsize CU--TP--914}\\
\vspace*{2.0cm}
\end{flushright}
Dynamical QCD thermodynamics with domain wall fermions}
 
\author{P.~Chen,
N.~Christ,
G.~Fleming,
A.~Kaehler,
C.~Malureanu,
R.~Mawhinney,
G.~Siegert
\thanks{Supported by the Max Kade foundation.},
C.~Sui, \ \ \ 
P.~Vranas
\thanks{Talk presented by P. Vranas. Current address: Physics Dept., University of Illinois, Urbana, IL 61801. Work supported in part by the Department of Energy.},
and
Y.~Zhestkov
\address{Physics Dept., Columbia University, New York, NY 10027}
}

\begin{document}

\begin{abstract}

We present results from numerical simulations of full, two flavor QCD
thermodynamics at $N_t=4$ with domain wall fermions. For the first
time a numerical simulation of the full QCD phase transition displays
a low temperature phase with spontaneous chiral symmetry breaking but
intact flavor symmetry and a high temperature phase with the full
$SU(2) \times SU(2)$ chiral flavor symmetry.

\end{abstract}

\maketitle

\section{Introduction}
\label{sec:intro}

QCD thermodynamics has been extensively studied using numerical
simulations with staggered or Wilson fermions. However, both
formulations break the chiral symmetry of the theory. The symmetry is
recovered together with the Lorentz symmetry as the continuum limit is
approached. In the past few years a novel fermion regulator was
developed that provides a way of controlling the amount of chiral
symmetry breaking at any lattice spacing.  Domain wall fermions
\cite{Kaplan} utilize an extra space--time dimension with free boundary
conditions to separate the two chiral components of the Dirac spinor.
The components are localized on the opposite boundaries (walls). If
the extent of the fifth ``dimension'', $L_s$, is infinite the two
chiral components decouple and the theory has the full chiral
symmetry, even at finite lattice spacing. For practical numerical
simulations, $L_s$ must be finite and as a result there is a small
mixing of the chiral components resulting in a residual mass. The
important point is that the size of this residual chiral symmetry
breaking can be controlled at any lattice spacing by increasing
$L_s$. For the first time the approach to the chiral limit has been
separated from the approach to the continuum limit.

In free field theory the localization is exponential and the
effective quark mass is given by \cite{PMV}:
\begin{equation}
\label{eq:free_m_eff}
m_{eff}^{(free)} = m_0(2-m_0) \left[m_f+(1-m_0)^{L_s}\right] .
\end{equation}
where $0 \leq m_0 < 2$ is the domain wall ``height'' and $m_f$ an
explicit fermion mass. In the interacting theory there is numerical
non-perturbative evidence that the general features of
eq. \ref{eq:free_m_eff} still hold with $m_0$ being renormalized \cite{RDM},
\cite{Fleming}, \cite{DWF_Columbia_BNL}.

Another unique property of domain wall fermions is that the
$L_s=\infty$ limit can be studied using the overlap formalism
\cite{NN1}. In that limit it can be shown that for topological gauge
field backgrounds there are exact fermionic zero modes.  It has been
demonstrated that for masses in the range of interest the effects of
finite $L_s$ on the zero modes can be made arbitrarily small for
classical \cite{CU_zero_modes}, \cite{Adrian} and for quantum
\cite{Fleming}, \cite{Lagae} gauge field backgrounds.  Domain wall
fermions possess the key ingredient in reproducing anomalous effects.

For more details on domain wall fermions see the review \cite{DWF_reviews}
and references within.

\section{Numerical Simulations}
\label{sec:hmc}

Dynamical domain wall simulations have been done in the past for the
Schwinger model in the $L_s=\infty$ limit (overlap formalism)
\cite{NNV} and for finite $L_s$ \cite{PMV} using standard hybrid Monte
Carlo (HMC) techniques. In this work we simulated QCD using standard
HMC at finite $L_s$. The gauge fields are defined on the
four-dimensional lattice while the fermion fields on the
five-dimensional one. The five-dimensional Dirac operator is as in
\cite{Furman_Shamir} with even-odd preconditioning.  The bulk effects
of the $L_s$ heavy flavors were subtracted \cite{NN1} by introducing
five-dimensional bosonic fields as in \cite{PMV}.  We used the $\Phi$
algorithm with trajectory length of $0.5$ and step sizes $\sim 0.01 -
0.02$ resulting in acceptance rates $\sim 80 \%$.  The standard
conjugate gradient inversion algorithm (CG) was used with a typical
number of CG iterations ranging from $50 - 200$. We used initial
configurations in the opposite phase and $200-800$ thermalization
sweeps.  The computational cost is linear in $L_s$.

\section{The Phase Transition}
\label{sec:trans}

In order to investigate the feasibility of studying QCD thermodynamics with
domain wall fermions, to locate the critical coupling, and to
investigate the parameter space of this new regulator ($m_0, m_f,
L_s$) we performed a large number of simulations at various couplings 
on $8^3 \times 4$ and $16^3\times 4$ lattices using the {\it QCDSP}
machine. This was possible due to the robustness of the machine and
software that allowed us to split a 200 Gflops portion of it to 7
independent, 6 Gflops machines and 6 independent, 25 Gflops machines. 

A sharp change on the value of the chiral condensate $\pbp$ and the
magnitude of the Wilson line $|W|$ was observed as $\beta = 6 / g_0^2$
(with $g_0$ being the gauge coupling constant) was varied. This can be
seen in figure 1 for an $8^3\times4$ lattice with $m_0=1.9$, $m_f=0.1$
and $L_s=12$. The fits are to $c_0 [c_1 + \tanh(c_2 [\beta_c - \beta])]$.
In the broken phase we did a dynamical mass
extrapolation of $\pbp$ at $\beta=5.2$ with $L_s=16$ and found a non
zero intercept, $9.1(1) \times 10^{-3}$, indicating the expected spontaneous
breaking of chiral symmetry. 
In the symmetric phase a dynamical mass
extrapolation of $\pbp$ at $\beta=5.45$ with $L_s=8$ resulted in an
intercept of $2.8(1) \times 10^{-3}$ but when $L_s$ was increased to
$L_s=16$ the intercept dropped to $4.5(1.3) \times 10^{-4}$ indicating
that the full chiral symmetry was restored to a high degree.
Similar results were obtained on a $16^4\times 4$ lattice at
$\beta=5.45$, $m_0=1.9$ and $L_s=16$, resulting in an intercept of
$4.4(5) \times 10^{-4}$. The dynamical mass data and extrapolations are
presented in figure 2.

The $L_s$ dependence of our results is presented in figure 3.  The
lines are curves of the form $c_0 + c_1 e^{- c_2 L_s}$ that go through
the three points.  The horizontal lines are the $\pbp=c_0$ lines. As
can be seen in the broken phase ($\beta=5.2$) where the coupling is
larger, the decay rate $c_2$ is larger than the one in the symmetric
phase ($\beta=5.45$) as expected from the Schwinger model \cite{PMV}.
At $L_s=16$ and $\beta=5.2$ $\pbp$ is within $7\%$ of the extrapolated
value while at $\beta=5.45$ it agrees within the error bars.

\fig{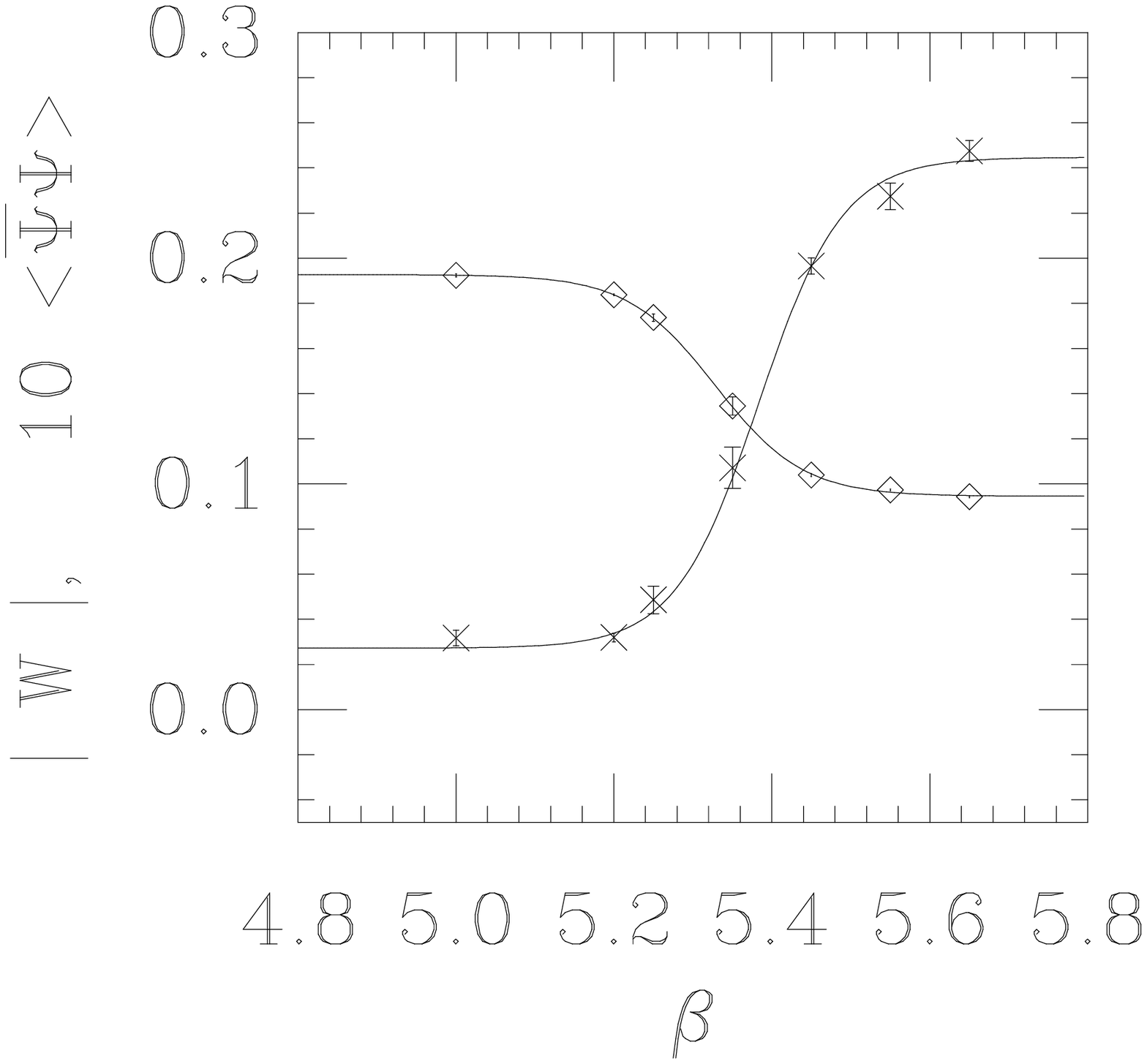}{\one}{\figsizeA}

\fig{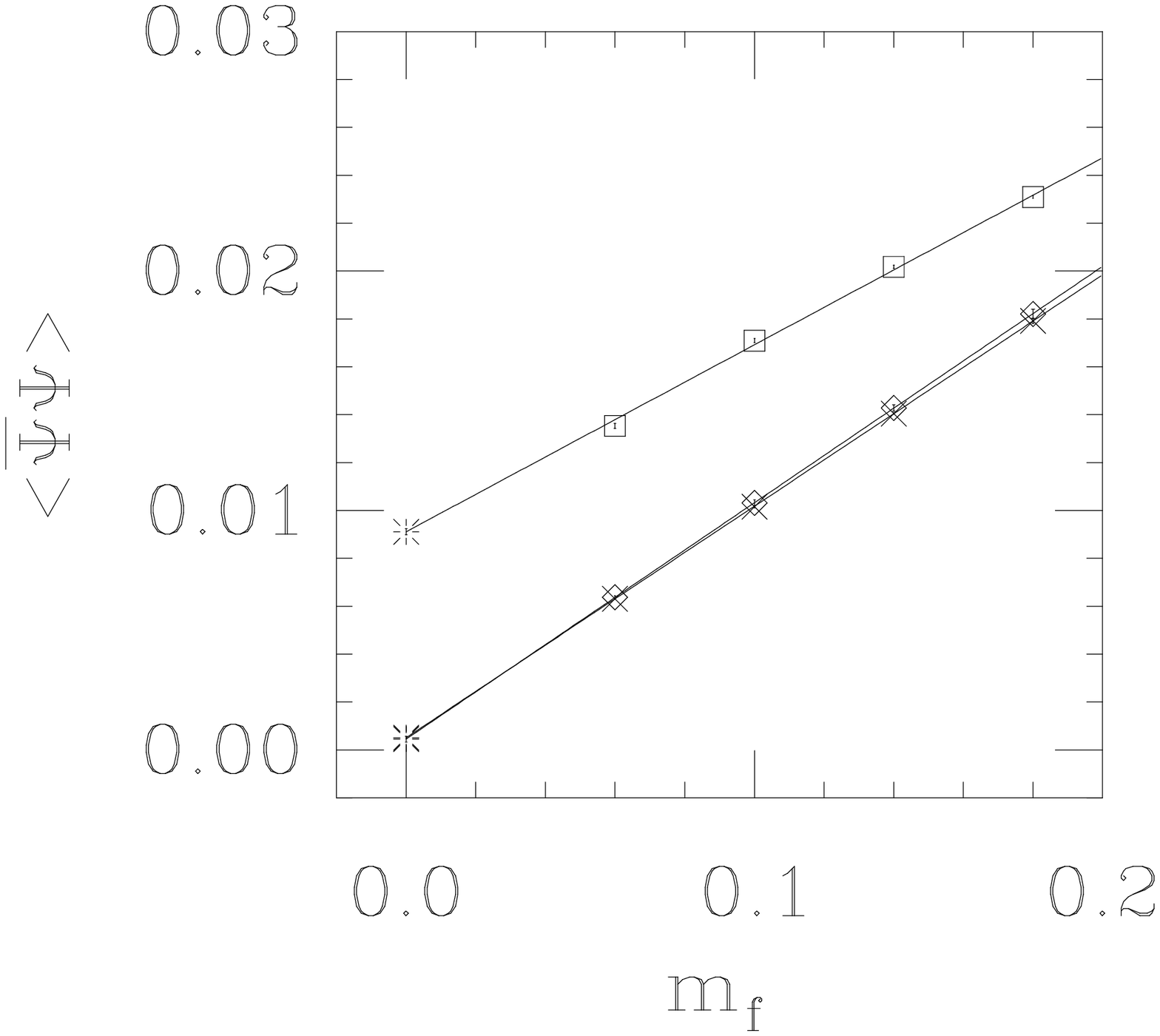}{\two}{\figsizeC}

The dependence on $m_0$ is more subtle.  In free
field theory $m_0 < 0$ corresponds to no light flavors, $0 \leq m_0 < 2$ to
one and $2 \leq m_0 < 4$ to four light flavors. Also, for positive but
small $m_0$ the number of low momentum states becomes small (for more
details and numerical results see \cite{RDM}). In
order to study these effects we located $\beta_c$ using
$\pbp$ and $|W|$, for various values of $m_0$ on $8^3 \times 4$
lattices. The results are in figure 4. The dependence of
$\beta_c$ on $m_0$ can be viewed as an expected renormalization
effect.

\fig{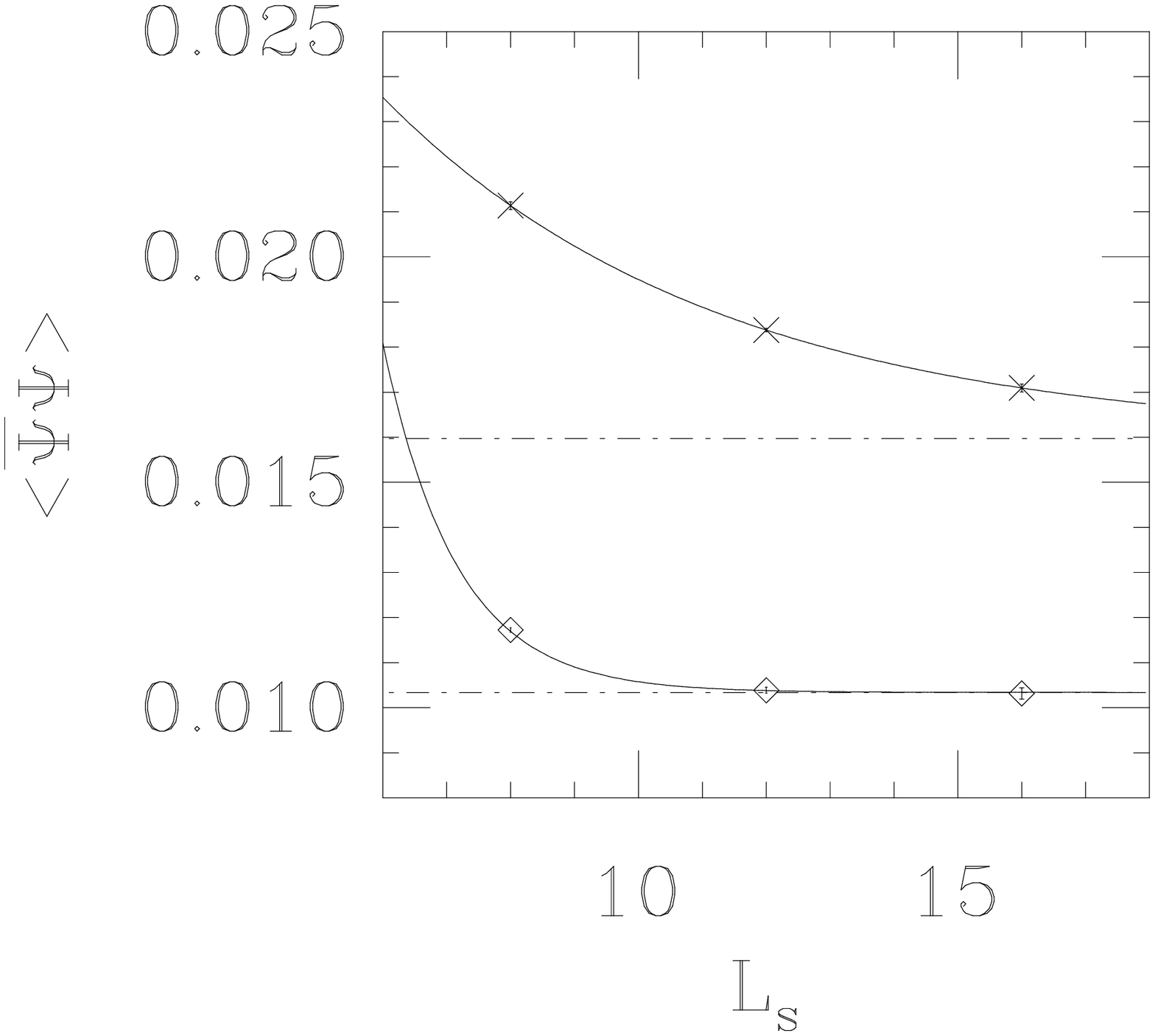}{\three}{\figsizeA}

\fig{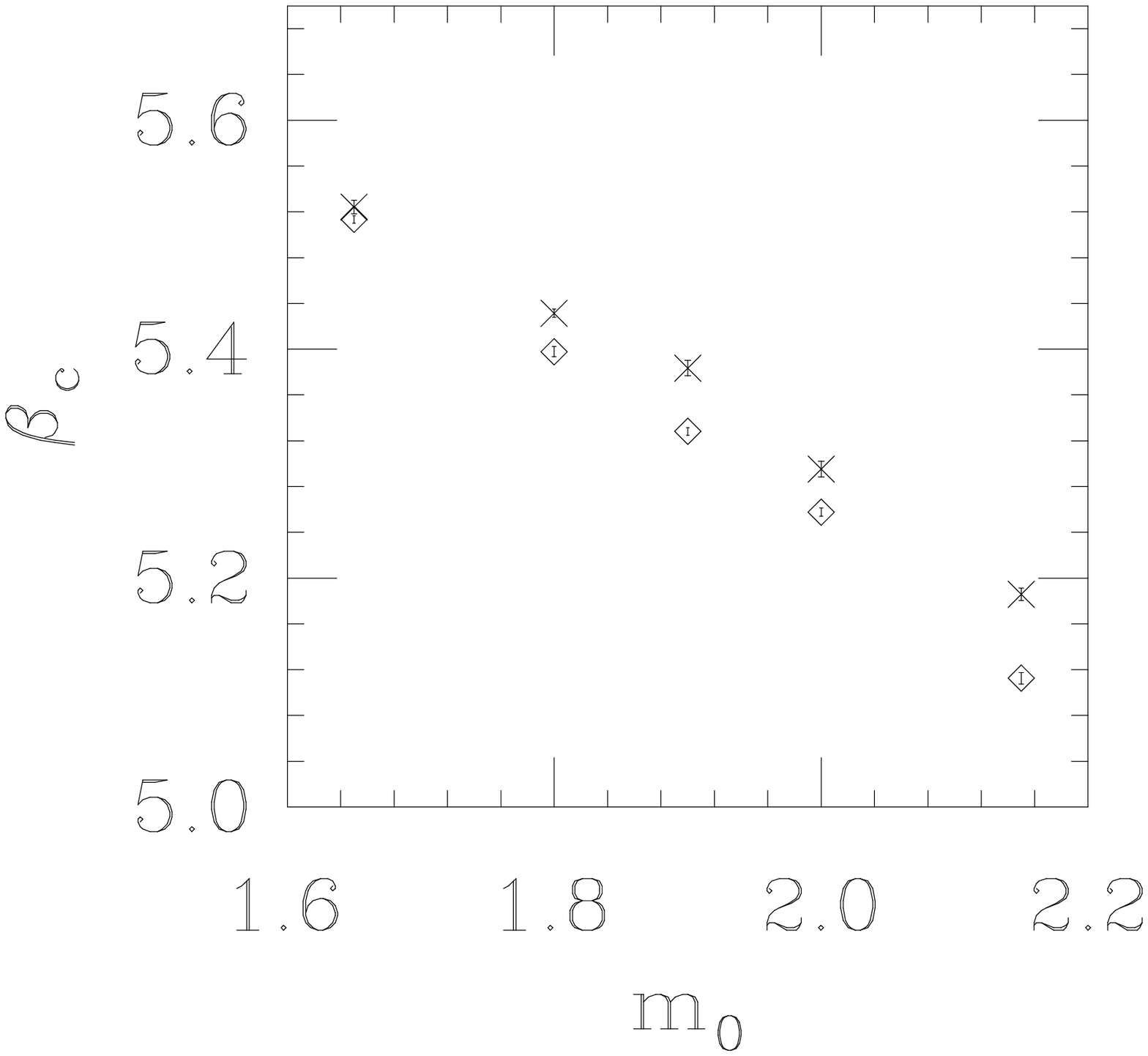}{\four}{\figsizeA}

\section{The $U(1)$ Axial Symmetry}
\label{sec:u1}

Due to the exceptional zero mode properties of domain wall fermions
the possibly anomalous breaking of the $U_A(1)$ symmetry just above
the transition can be investigated with clarity and without subtleties
such as zero mode shift effects (staggered fermions) or proximity to
the Aoki phase (Wilson fermions).  We measured the difference of the
screening masses of the $\delta$ and $\pi$ as a function of $m_f$ on a
$16\times 4$ lattice at $\beta=5.45$ with $m_0=1.9$ and $L_s=16$. The
data fits to $c_0 + c_2 m_f^2$ with $\chi^2/dof \approx 0.5$ and $c_0
= 3.1(9) \times 10^{-2}$.  The small $\chi^2/dof$ and the near-zero,
small-mass limit of $\pbp$ indicate that the chiral symmetry breaking
effects due to the fermion regulator ($L_s < \infty$) are negligible
so that this preliminary, non-zero $c_0$ value suggests that $U_A(1)$
remains broken just above the transition.

\fig{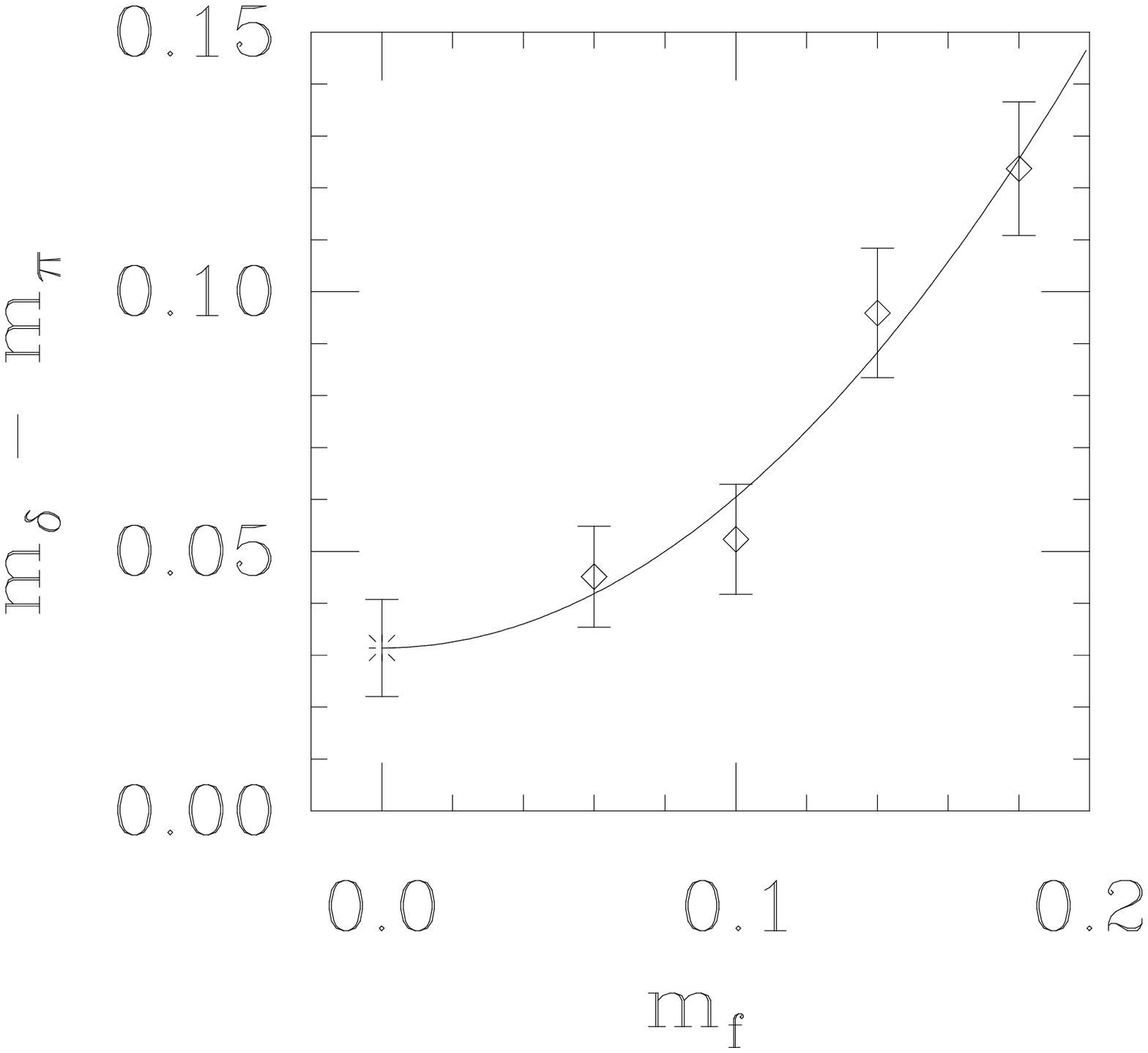}{\five}{\figsizeB}

\section{Conclusions}
\label{sec:conclusions}

We presented results from numerical simulations of full, two flavor
QCD thermodynamics at $N_t=4$ with domain wall fermions. They indicate
the presence of a low temperature phase with spontaneous chiral
symmetry breaking but intact flavor symmetry and a high temperature
phase with the full $SU(2) \times SU(2)$ chiral flavor symmetry.
Given this, we interpret the difference seen in the $\pi$ and
$\delta$ screening lengths just above the transition as preliminary
evidence for the anomalous breaking of the $U_A(1)$ symmetry.

\end{document}